\documentclass[amsmath,amsfonts,aps,prb,twocolumn,showpacs]{revtex4}

\usepackage{graphicx}
\usepackage{dcolumn}
\usepackage{bm}

\begin{document}
\preprint{OST}

\title{Optical spin transfer in ferromagnetic semiconductors}

\author{J. Fern\'andez-Rossier$^{1,2}$, A. S. N\'u\~nez$^{1}$, 
M. Abolfath$^{1,3}$, and A. H. MacDonald$^{1}$}
\affiliation{(1) Physics Department, University of Texas at  Austin. 
 1 University Station C1600, Austin,  TX 78712 \\ 
 (2) Departamento de F\'isica Aplicada, Universidad de Alicante. San Vicente del
 Raspeig 03690, Spain \\
 (3)  Institute for Microstructural Sciences
National Research Council of Canada
1200 Montreal Road
Ottawa, Ontario K1A 0R6}

\date{\today} 

\begin{abstract}

Circularly polarized laser pulses that excite   electron-hole pairs across the
band gap of (III,Mn)V ferromagnetic semiconductors can be used to  manipulate
and to study collective  magnetization dynamics. 
The initial spin orientation of
a photocarrier in a (III,V) semiconductors is determined by the polarization state
of the laser.  We show that the photocarrier spin can be irreversibly
transferred to the collective magnetization, whose dynamics  can consequently
be flexibly controlled by suitably chosen laser pulses.  As illustrations we
demonstrate the feasibility of all optical ferromagnetic resonance and
optical magnetization reorientation.
\end{abstract} 

\maketitle
  
\section{Introduction}

The collective magnetization dynamics of a single-domain ferromagnet 
can be dramatically modified when spin polarized quasiparticles are
injected into the system. For instance, in a metallic ferromagnet 
current-carrying, non-equilibrium quasiparticles exert a torque on
the collective magnetization which, at sufficiently high current densities, can
produce a complete reversal of the magnetization orientation. This phenomenon,
called spin transfer (ST), was  predicted by Slonczewski \cite{SMT1}
and  Berger \cite{SMT2} and has been confirmed experimentally in 
ferromagnetic multilayer systems by a number of groups \cite{SMTexp}.
Although the microscopic mechanism is not completely settled and possibly not
absolutely universal, it is clear that ST in itinerant electron ferromagnets 
is a consequence of irreversible transfer of magnetization between 
non-equilibrium quasiparticles and the collective magnetization.
In ST, a spin-polarized injection current provides 
a non-conservative driving force which can either deliver or extract energy 
from the collective magnetic degree of freedom.

(III,Mn)V ferromagnetic semiconductors like GaAs:Mn  combine ferromagnetism
with familiar semiconductor properties similar to those of the parent
semiconductor \cite{OhnoScience}.  Most practical applications of GaAs and other
III-V compounds are related to their optical properties. Unlike Si and Ge,
III-V materials are optically active and therefore   respond strongly to a
laser field with a frequency close to the band gap.  A well known property of
III-V semiconductors is optical orientation \cite{OO1} in which a laser
generates a population of photocarriers strongly {\em spin polarized} along a direction
which depends on the polarization state of the laser field.   In this paper we
predict that circularly polarized laser pulses which  excite spin polarized
electron-hole pairs across the bandgap of (III,Mn)V ferromagnets can control
the magnetization dynamics through the  spin transfer phenomenon.  We outline a
theory of optical spin transfer in ferromagnetic semiconductors and discuss
some of  the many possible applications of this phenomenon.  In particular, by
numerically  solving Landau-Lifshitz equations that include a spin transfer
term, we show that laser pulses with suitably chosen durations, intensities, and
propagation directions enable all-optical ferromagnetic resonance, and
nanosecond time scale switching between magnetic easy axes. 

The rest of this paper is organized as follows. In section II we briefly review
aspects of the electronic structure of (Ga,Mn)As  relevant to our
proposal.  In section III we present our  theory of optical spin transfer,
which adds to the Landau Lifshitz equations (LLE) that describe collective
magnetization dynamic an additional term that accounts for the
irreversible exchange of angular momentum
with the photocarriers.  In section IV we present results of the numerical
solution of the extended LLE in a number of different cases. In section V we
discuss the relationship between our proposal and some recent experimental results
\cite{Munekata,Munekata-03}.  We have relegated a description of some of the 
technical considerations that underly our theory to appendices A and B.

\begin{figure}[hbt]
\includegraphics[width=7cm]{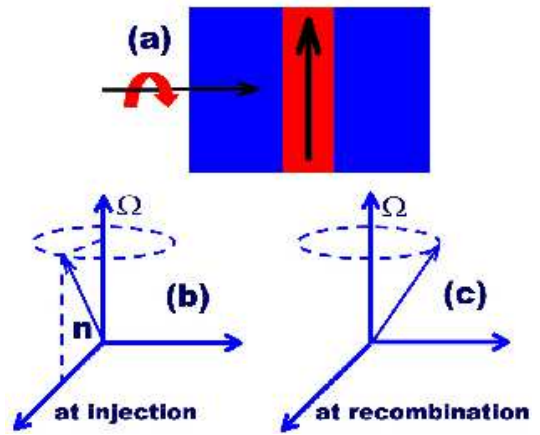}
\caption{\label{fig1} Schematic summary of optical spin transfer theory}
\end{figure}

\section{ Electronic structure of (Ga,Mn)As}

The Fermi energy of intrinsic GaAs lies in the middle of  a gap of 
approximately 1.5 eV. GaAs can be
doped with up to 10 percent of Mn which, under suitable growth conditions,
replaces Ga. Substitutional Mn acts as a relatively shallow single acceptor so
that each Mn injects one hole in the GaAs host. In most of the samples there is
some degree of compensation, due primarily to interstitial Mn ions, so that the hole density is smaller than
density of magnetic impurities.  Samples in which the density of interstitial Mn ions 
has been reduced by annealing tend to have higher carrier densities and 
metallic transport properties, indicating that the
holes are delocalized.  These relatively itinerant holes interact via a 
Kondo-like interaction with the localized $d$ orbitals of the Mn ions. 
There is broad agreement that ferromagnetism is due to carrier mediated
interactions between the Mn ions. Strong magneto-transport effects
\cite{magnetotransport} and tunneling magnetoresistance \cite{TMR} 
support this point of view and demonstrate that the valence band holes are strongly coupled to the
Mn ions.

The physics of samples with metallic-like conductivity can be reasonably
\cite{note}
described with a simple mean field theory: the
valence band of the parent compound (GaAs), described with a
$\vec{k}\cdot\vec{p}$ Hamiltonian, is occupied by 
holes which interact with an effective field that is the mean
field representation of the exchange interaction with the array of Mn atoms
\cite{Zener,Ramin,JFR}. 
Ferromagnetism occurs at low temperatures, when the paramagnetic energy gained
by the degenerate Fermi gas formed by the holes is larger than the reduction of
entropy of the magnetic atoms entailed by the spontaneous order. 
Importantly, the spin orbit interaction is properly included in the
$\vec{k}\cdot\vec{p}$
Hamiltonian. As a result of spin orbit interaction, both the 
$\vec{k}\cdot\vec{p}$ valence bands $\epsilon_{\nu}(\vec{k})$ and the total energy
$\cal{E}\left({\cal \vec{M}}\right)$ 
depend on the relative orientation of the collective magnetization
$\cal{\vec{M}}$ and the crystallographic axis. This magnetic anisotropy 
 compares well with  experimetal results \cite{Tang}.

The conduction band does not play an important role in equilibrium
(Ga,Mn)As  and remains empty unless electrons are generated there 
by optical excitation, as is the case of our proposal.
We describe the bottom of the conduction band with a simple parabolic band model. The mean field
spin splitting is given by $\Delta_c \equiv J_{sd} c_{Mn} $, where $J_{sd}$ is
the exchange coupling between the conduction band electrons and the Mn $d$
electrons and $c_{Mn}$ is the density of Mn ions.
To the best of our knowledge, the conduction band exchange coupling
constant $J_{sd}$ has not been measured experimentally in (Ga,Mn)As. 
In analogy with the case of
(II,Mn)VI  \cite{ReviewFurdyna}, we {\em assume} that  $J_{sd}$ is five times
smaller than $J_{pd}$, the exchange interaction between valence band holes and Mn ions.
The latter can be inferred from transport and magneto-optics experiments. We
take $J_{sd}= 11 eV$ $\AA^3$. (Our conclusions do not depend strongly on the  
numerical value of $J_{sd}$.)

\section{ Optical Spin Transfer} 

In equilibrium, the  magnetization $\cal{\vec{M}}$ of 
a sample of (Ga,Mn)As with a density $p$ of holes,
  lies along  an easy axis in order to
minimize $\cal{E}\left({\cal \vec{M}}\right)$. In this paper we study the
dynamics of $\cal{\vec{M}}$ when the  material is photoexcited so that a 
density of extra holes, $\delta p$ and extra conduction band electrons $\delta
n=\delta p$ is injected in the system. We only consider the situation 
when $\delta n= \delta p<<p$. The initial spin orientation of these
photocarriers, $\hat{n}$,  is determined by the polarization state of the laser, according to
the selection rules of the material which depend ultimately on an interplay between
angular momentum conservation and spin-orbit interactions. A circularly
polarized laser propagating along the $\hat{z}$-direction with an  
energy equal to the band-gap creates photocarriers that are strongly spin polarized along the propagation
direction \cite{OO1}.

The Mn spin dynamics  of (III,Mn)V ferromagnetic
semiconductors differs qualitatively from that  of the paramagnetic (II,Mn)VI
semiconductors which have been studied extensively \cite{DMSOO,Crooker} in
interesting earlier work.  In that case, laser pulses have been used both to
trigger and to detect \cite{DMSOO,Crooker} the dynamics of substantially
independent magnetic moments. In ferromagnets, moments behave collectively and
many elegant and technologically important properties follow from the, often
complex, behavior of the magnetization-orientation collective coordinate.  In
the case of (III,Mn)V ferromagnets \cite{ourreview} the underlying magnetic
degrees of freedom are Mn ion  $S=5/2$ local moments and holes in the
semiconductor valence band \cite{note}. The magnetization-orientation dynamics
is governed by the following equation:
\begin{equation}
\frac{\partial\vec{\cal M}}{\partial t}= 
\vec{\cal M} \times \left[ -\gamma \frac{\partial {\cal
E}(\vec{\cal M})}{\partial \vec{\cal M}}
+ \vec{\Gamma}_{\rm damping} + \vec{\Gamma}_{\rm ST} \right]
\label{LLS}
\end{equation}
where ${\cal E}(\vec{\cal M})$ specifies the relationship between energy 
and magnetization, $\vec{\cal M}$. The dissipative processes by which the collective coordinate relaxes
towards the minimum of ${\cal E}(\vec{\cal M})$ are represented by $\vec{\Gamma}_{\rm
damping}$.  Choosing the Landau-Lifshitz form, the damping term is given by
$$\vec{\Gamma}_{\rm damping}=-\frac{\gamma\alpha}{{\cal M}_s} \vec{\cal M}\times \frac{\delta {\cal
E}(\vec{ \cal M})}{\delta\vec{\cal M}}$$
 where ${\cal M}_s = |\vec{\cal M}|$
and $\gamma= \frac{e}{m c}$ is the gyromagnetic ratio.  We define the unit
vector $\vec{\Omega}\equiv \frac{\vec{\cal M}}{{\cal M}_s}$; note that since $d ({\vec{\cal M}} \cdot
{\vec{\cal M}})/dt =0$, these equations attempt to describe only the dynamics of the 
magnetization orientation ${\vec{\Omega}}$.

In this work the irreversible transfer of angular momentum and energy from 
non-equilibrium quasiparticles to the collective magnetization is described by
$\vec{\Gamma}_{\rm ST}$.  The expression that we use for $\vec{\Gamma}_{\rm
ST}$ is based on the following physical picture.  After photon absorption, the 
photocarriers are spin polarized along the direction $\hat n$ determined by the
polarization state and the  propagation direction of the laser (figure 1-b).
Due to their mutual exchange interaction \cite{ourreview}, the spin of a photocarrier
and the collective magnetization will then precess around each other at rates 
defined by their mean-field interactions.  The precession involves periodic
{\em reversible} transfer of angular momentum back and forth between the
collective magnetization and an individual photocarriers, which can be described either
classically or quantum mechanically. The precession is abruptly interrupted by
one of the following two processes: spontaneous emission of a photon
(photocarrier recombination) or photocarrier spin decoherence. As we discuss
below, the latter process will usually involve spontaneous emission of a spin
wave. Importantly, the distribution of values for the component of the
photocarrier spin {\em perpendicular} to $\vec{\cal M}$, $\hat{n}_{\perp}$,
is uniform at the instant of decay, provided that the decay time is much
larger than the precession time (see figure 1-c). Therefore, when summing over
all the photocarriers, there is a {\em net loss of quasiparticle spin}
along the $\hat{n}_{\perp}$ direction {\em which  must be transferred to the collective
magnetization} because of the  conservation of total spin during precession
\cite{SMT1}. Therefore, the stochastic nature of the spin decoherence event,
whether due to carrier recombination or spin wave emission,  leads to an
irreversible spin transfer from the photocarriers to the collective
magnetization.  This argument implies that:
\begin{equation}
\vec{\Gamma}_{\rm ST}=  \frac{\mu_B {\cal P}(t)}{{\cal M}_s^2}
\vec{n}\times\vec{\cal M}
\label{ST}
\end{equation}
where ${\cal P}(t)$ is the rate per unit volume at which quasiparticles
are injected into the system and $\mu_B$ is the Bohr magneton. 
In appendix A we present a
 mathematical derivation of eq. (\ref{ST}).  
A  term similar to eq. (\ref{ST})
has been proposed previously on the basis of similar arguments
in the context of current induced magnetization switching in
ferromagnetic metals \cite{SMT1,SMT2}.

%\subsection{Photocarrier spin relaxation}
The argument leading to Eq. (\ref{ST}) assumes that the photocarriers precess
many times around the effective field created by the  collective magnetization
before they loose spin coherence.   The major source of spin decoherence in the
case of the electrons in the conduction band in non magnetic p-doped GaAs  at
low temperatures is exchange with holes \cite{BAP}. The measured
\cite{electronrelax} electron spin relaxation time in p-doped samples is well
above 100 ps at low temperatures. To the best of our knowledge, the conduction
band electron spin relaxation rate has not been measured in(Ga,Mn)As.
Using the standard master equation approach with Fermi Golden rule rates
we have derived an expression for 
the conduction band spin decoherence time. The details are outlined  
in appendix B. Whereas the static component of
the magnetic environment produces the conduction band spin splitting, $\Delta_c$, 
the fluctuating magnetic environment provided by
the spin waves of the Mn-hole system results in the following $T_2$: 
 \begin{equation}
\frac{1}{T_2}=\left(1+e^{-\beta_L \Delta_c}\right)
 S \; \frac{J_{sd}^2(k_B T_e)^2}{4 \pi^2\hbar} 
   \left[\frac{2m }{\pi\hbar^2\cal D}\right]^{3/2} 
% \times \nonumber \\
%&\times&
{\cal F}(y)
 \label{T2}
 \end{equation}
 where
 \begin{eqnarray}
{\cal F}(y)=  \int_{0}^{\infty}   \sqrt{x}
 e^{-x} \int_{0}^{z_D}  
 \sqrt{x+y-\frac{T_e}{T_L}z}\times
 \nonumber \\ \times  
 \sqrt{\frac{T_e}{T_L}z}\;
 \left[1+n_B\left(\frac{T_e}{T_L}z\right)\right] dzdx
 \label{F}
\end{eqnarray}
In this expression we distinguish between the Mn temperature $T_L$ and the
temperature of the photocarriers, $T_e$, which is taken as a parameter to
describe their excess energy. Here ${\cal D}\equiv \frac{2A}{c_{Mn}S}$ is the
spin stiffness \cite{SW} in the spin wave
spectrum  $\Omega=  {\cal D}q^2$, $m^*$ is the conduction band effective mass,  $z_D=
\beta k_B^2\left(6 \pi^2 c_{Mn}\right)^{2/3}$ is the normalized spin wave Debye
cutoff \cite{Konig} and $y=\Delta_c/T_e$. Since  and $y>>z$ 
the argument of the square root in eq. (\ref{F}) is always positive.
The physical mechanisms underlying
equation (\ref{T2}) are the spontaneous emission and absorption of spin waves
with the corresponding spin flip of the conduction band electron. 
Since the spin wave gap is neglected in the derivation of
equation (\ref{T2}), $T_2$ is probably somewhat underestimated. 
$T_2$ is a decreasing function of the hot carrier temperature $T_e$. 
If we take $S=2.5$,  $c_{Mn}=$1.1$\times10^{21}$ cm$^{-3}$ (x=0.05),
$A= 0.2 pJ$ $m^{-1}$ (according to reference \onlinecite{SW}), 
$m^*=0.067$,$T_L=1 meV$ and  $J_{\rm sd}=$ 11 eV $\AA^3$, then
we have $T_2\simeq20$ ps for $T_e=$ 100 meV, and even longer $T_2$ for smaller
values of $T_e$.   
 This  time is long compared to
the precession period for photoelectrons which is  $\sim h/ J_{sd} S c_{\rm Mn}
\sim 0.15$ ps, where $c_{\rm Mn}$ is the Mn concentration, and $J_{sd}$ is the
exchange interaction between Mn moments and conduction band electrons.  The
photocarrier spin-orientation randomization assumption that underlies Eq.
(\ref{ST}) is therefore valid for electrons in the conduction band.  Radiative
recombination time, by which one photo-electron and one photo-hole annihilate
by emission of a photon, lies in the range  between  2   and 25 ps 
\cite{Munekata,Munekata-03}. We expect that both   carrier recombination and spin wave
emission will contribute significantly to the spin decay of photoelectrons.

Unlike photoelectrons, photoholes experience a strong momentum-dependent 
effective field $\frac{1}{3} \Delta_{so} \vec{L} $, due to their spin-orbit interaction with the
atomic orbital angular momentum, $\vec{L}$. In GaAs, for example, $\Delta_{\rm
so}=340$ meV, larger than the valence band exchange mean field, $\Delta_{\rm
pd} = SJ_{pd}c_{\rm Mn}$. Scattering between Bloch states causes $\vec{L}$ to
fluctuate strongly  so that the valence band spin decoherence time is on the
same order  as the momentum scattering time \cite{EY}. In (Ga,Mn)As the
momentum scattering time is of the order of 10 femtoseconds \cite{conduc},
shorter than or comparable to  the band quasiparticle spin-precession time
$\sim h/\Delta_{\rm pd} \sim 30 \,\,{\rm fs}$.  These numerical values in 
combination with our picture of optical spin transfer suggest that the
contribution of photo-holes will be strongly suppressed.  We neglect 
this contribution in the rest of the paper. The effect of spin
orbit interaction in spin transfer has been studied theoretically by two of us
\cite{alvaro}, for a model without orbital degeneracy. 
We believe that this issue calls for further work, especially in the
light of recent experiments in which current driven magnetization switching is
achieved at a much smaller than expected current density\cite{Ohno2004}.  

\section{Optical Spin Transfer Dynamics in (Ga,Mn)As} 
In the remainder of this paper we explore the  collective magnetization
dynamics of (Ga,Mn)As driven by polarized laser pulses,  as described by Landau
Lifshitz equations that include the spin transfer term (eq. \ref{LLS}). 
In the model of valance-band-hole mediated 
ferromagnetism \cite{Zener,Ramin}, the magnetic anisotropy is due to the spin orbit
interaction of the holes and is sensitive to lattice mismatch strains.
The anisotropy energies we consider can be fit to the form
%\begin{eqnarray}
$${\cal E}(\vec{\Omega})=
%&=&
 k_1 \Omega_x^2 \Omega_y^2 + k_2 \Omega_z^2  -k_3
\Omega_z^4  
%+\nonumber \\&+&
+k_4 \left( \Omega_x \Omega_y \Omega_z\right)^2  +k_u \Omega_x^2
$$
%\end{eqnarray}
The uniaxial term proportional to $k_2$ is a consequence of lattice-matching strains and favors magnetization
orientations in the $\hat x-\hat y$ plane,  while the much weaker uniaxial term proportional to $k_u$ is
obtained by fitting to recent experiments \cite{Tang} which demonstrate that the kinematics of  the MBE
growth process lowers the symmetry of the layer-by-layer growth planes in a way which influences the magnetic
anisotropy energy.  The following set of values $k_1=0.025 $ meV/nm$^3$,  ($k_2,k_3,k_4,k_u)=1.34,1.05,
-1.25,0.08)$ in units of $k_1$ follow from the numerical mean field calculation \cite{Ramin}, for 
(Ga$_{0.95}$,Mn$_{0.05}$)As with a density of holes $p=3.5\;10^{20}$ cm$^{-3}$.
 These values are in good
agreement with the experimental results reported by Tang {\em et al.} \cite{Tang} and can be taken as typical
for GaAsMn. We have verified that the results reported below are robust with respect to modifications  in the
functional ${\cal E}(\vec{\Omega})$. 

Writing ${\cal E}(\vec{\Omega})=k_1 e_0(\vec{\Omega})$, where $e_0$ is
dimensionless, we define a typical time scale 
$t_0\equiv \left|\frac{{\cal M}_s}{\gamma k_1}\right|\sim 30$ $ps$ for 
a Mn fraction x=0.05. Defining the dimensionless quantities
$\hat{t}=\frac{t}{t_0}$ and  $\hat{\cal P}\equiv  \frac{ {\cal
P}(\hat{t}) t_0}{S c_{\rm Mn}} $, 
the Landau Lifshitz equation can then be written as
\begin{equation}
\frac{d\vec\Omega}{d\hat{t}}= \vec{\Omega} 
\times \left[- \frac{\partial  e_0}{\partial \vec{\Omega}}
-\alpha \left[ \vec{\Omega} 
\times \frac{\partial e_0}{\partial \vec{\Omega}} \right]
+ \hat{\cal P}(\hat{t}) \left[\hat{n} \times \vec{\Omega} \right] \right]
\label{LLS2}
\end{equation}
where the damping coefficient $\alpha$,  estimated
using the ferromagnetic resonance experiments of reference \cite{FMR}, is
  $\alpha\simeq 0.07$.  We use this value for the calculations that
follow, although our conclusions are not particularly sensitive to this
parameter. We  assume that the magnetization lies along the
$\vec{\Omega}_0=\hat{x}$ easy axis before the laser pulse is applied and
consider two different situations: {\em i)} Weak circularly polarized laser
pulses which propagate along $\hat z$ and  initiate a free induction decay from
which a ferromagnetic resonance spectrum can be obtained without the need for a
time dependent magnetic field and {\em ii)} Intense circularly polarized laser
pulses, propagating either parallel or perpendicular to $\vec{\Omega}_0$ which
drive $\vec{\Omega}$ away from $\hat{x}$ and switch the magnetization
direction. 

\subsection{ All optical Ferromagnetic Resonance} 
We first discuss the possibility of performing a time resolved pump and probe
experiment which yields information similar to that of a ferromagnetic
resonance (FMR) experiment. We refer to this as all optical FMR.  In standard
FMR experiments, the dynamics of the magnetization is triggered by an AC 
magnetic field, whereas  detection  of the magnetization dynamics can be done
using different methods, including optical ones.  Here, we show that the motion of
the collective magnetization can be both triggered and detected with  pump
and  probe laser pulses. The pump laser, propagating in the $\hat
z$ direction, is circularly polarized, so that the photocarriers are spin
polarized perpendicular to the equilibrium magnetization ($\hat x$).  Because
of the spin transfer term, the magnetization can be tilted away from $\hat x$
(see inset of figure 2) by the pulse. 
The energy per pulse  considered in figure 2  is $E=0.1 mJ\;cm^{-2}$ and the
laser duration is 3$ps$, corresponding to a laser power of 
$W=33 MW\;cm^{-2}$. Very similar results are obtained for pump pulse 
widths between $0.2$ and $20$ picoseconds, keeping the laser power constant:
it is primarily the laser power that 
controls the the spin transfer effect on the magnetization,
as implied by equation (\ref{ST}).

The subsequent free induction decay
process can be probed by measuring the  Faraday rotation of a second linearly
polarized probe laser pulse.  The Faraday rotation is proportional to $M_z(t)$,
which \cite{Crooker} oscillates at the  FMR frequency. In Fig. 2-b we show a 
normalized Fourier transform of $M_z(t)$ for two different values of the
damping constant, $\alpha$. It is apparent from Figure 2-b, both the precession
frequency, $\omega_0$  and the linewidth $\Delta \omega$,
 can obtained from such a procedure.

Although the experimental procedure we propose is similar to  the one   used in
reference \cite{Crooker} for diluted paramagnetic II-VI-Mn, the spin transfer
mechanism is quite different. In the case of II-VI-Mn quantum wells, the  spin
of the photoinduced heavy holes points along the growth direction, creating a
transient effective field $J_{pd} \;p_{\rm holes} \;\hat{z}$ which tilts the Mn
spins away from their equilibrium orientation \cite{Crooker,Linder}, determined
by an external magnetic field along $\hat x$. After recombination the transient
field is  absent and the Mn spins follow free induction decay dynamics.

\begin{figure}[hbt]
\includegraphics[width=6cm]{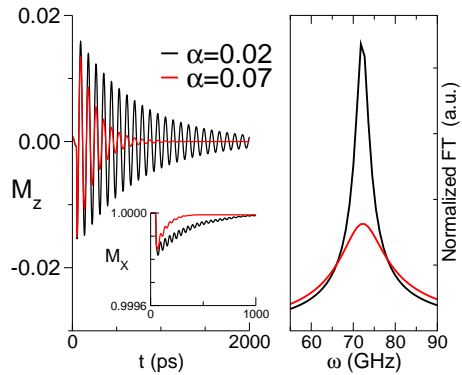}
\caption{ \label{fig2}
All optical FMR signal. The energy per laser pulse for the illustrated
simulation is $E=0.1 mJ\;cm^{-2}$ and the laser duration is $t_L=3$ ps. The
density of photocarriers corresponding those parameters and a extinction
coefficient $\alpha_L=10^{4}$ cm$^{-1}$ \cite{alpha} and recombination time
$t_R= 2$ps \cite{Munekata} is $1.2\times10^{18}$ cm$^{-3}$. Similar results are
obtained for pump pulse  widths between $0.2$ and $20$ picoseconds; it is
primarily the laser power that  determines the spin transfer effect on the
magnetization, as implied by equation (\ref{ST}).}
\end{figure}

\subsection{ Laser induced magnetization switching}
In the case of all optical FMR, the magnetization orientation is weakly
perturbed. By increasing the laser pulse intensity, the departure from the
initial equilibrium orientation can be made large enough to drive
$\vec{\Omega}$ to a {\em different} easy axis.  We consider two geometries the 
spin of the  photocarrier lying (a) perpendicular to $\vec{\Omega}_0$, and  (b)
parallel to  $\vec{\Omega}_0$. These situations were also
distinguished by Slonczewski \cite{SMT1}.  For that case the geometry (a) the
magnetization develops a precession around  the easy axis. In the geometry (b)
the effect of the spin transfer is to enhance or reduce the damping of the
departures of $\vec{\Omega}$ from $\vec{\Omega}_0$. For a sufficiently high
flux of non-equilibrium quasiparticles, the rate at which non-equilibrium
quasiparticles deliver energy into the collective magnetization can overcome
the rate at which the latter dissipates energy. When this happens,
$\vec{\Omega}$ departs from $\vec{\Omega}_0$ and,  depending on ${\cal
E}(\vec{\Omega})$, it will evolve to a different easy axis orientation.  We
have explored the two geometries using laser pulses. A  pulse that propagates
perpendicular to the magnetization drives $\vec{\Omega}$ from $\hat{x}$.  It is
possible to tailor the laser pulse energy and duration  so as to control  which
easy direction the magnetization decays to.  In particular, the laser pulse can
produce switching. In  figure 3 we show an example of this.  In the case  of
photocarriers injected with spin parallel to $\vec{\Omega}_0$, we have verified
that the laser can control the rate at which the collective magnetization
decays towards equilibrium. The damping can be made arbitrarily large, for
lasers propagating anti-parallel to the magnetization, or arbitrarily small for
lasers propagating parallel to the magnetization. 

\begin{figure}[hbt]
\includegraphics[width=5cm]{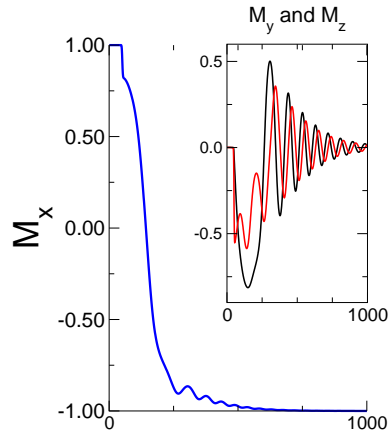}
\caption{ \label{fig4} Switching in the perpendicular configuration. Initially, 
the magnetization is pointing along $\hat x$. The sample is excited with a 
laser pulse of duration $t_L=3$ ps and energy density $E=4$ mJ/cm$^{-2}$. 
The density of photocarriers is $5 \times 10^{19}$ $cm^{-3}$. }
\end{figure}

\section{Discussion}

Ferromagnetic order of the microscopic degrees of freedom in (Ga,Mn)As,
 Mn $d$ electrons and valence band holes, is described with the 
 order parameter $\vec{\cal M}$.
  In equilibrium, $\vec{\cal M}$ lies
along some easy axis. An external perturbation can trigger the motion of the
order parameter, as described by the  Landau Lifshitz equations. In this
paper we describe a new type of external perturbation, the optical
injection of photocarriers taht are polarized along a direction $\hat n$
different from the orientation of the collective magnetization$\vec{\cal M}$. Our proposal takes
advantage of the selection rules for inter-band optical transitions in
(II,V) semiconductors. Provided that the
spin coherence time of the photocarriers is much longer than the precession time, we
predict a spin transfer (c.f. {\ref{ST}) term in the LL equation.
The spin-coherence time requirement is clearly met for conduction band electrons,
while the situation for valence band holes is less clear and calls
for further work. 
Numerical solutions of the LL equations with the spin transfer term show that
the magnetization dynamics in (Ga,Mn)As can be controlled with laser pulses. 

We now discuss the extent to which our theory can account for experiments recently
reported by Oiwa {\em et al.} \cite{Munekata-03}, in which the magnetization of
a GaAs:Mn film excited  by laser pulses of 120 $fs$ duration and power below 
5$\times 10^{12}$ photons/cm$^{2}$.  For a central photon energy of 1.579 eV,
the laser power per pulse is approaximately 6$MW$/cm$^{2}$, comparable with the
simulation of figure 2. Apart from the fact that both the experiment and our
theory describe the departure of the magnetization from the equilibrium
configuration induced by  spin polarized photocarriers, there is a number of
differences between the results of our model and those observed experimentally.
The Kerr rotation signal reported by Oiwa {\em et al.} \cite{Munekata-03} is an
exponentially decaying function with a time decay constant of less than 50 $ps$
in contrast with our figure 2, in which $M_z$ is an oscillating function whose
amplitude decays in a time scale of 2000 $ps$. The Kerr signal observed
experimentally is overdamped, the decay constant is smaller than the precession
period. In order to have an overdamped behaviour in our simulations, we would
need to take a Gilbert damping coefficient of $\alpha\simeq 5$, two orders of
magnitude larger than the value reported experimentally \cite{FMR}
At a more quantitative level, Oiwa {\em et al.} \cite{Munekata-03} claim that
each photo-hole is able to rotate 100 Mn spins in the case GaMnAs excited with
laser pulses and as much as 10$^6$ Mn spins in the case of cw experiments
\cite{Munekata}. In our theory angular momentum is exchanged between photocarriers
and Mn, and one photocarrier could not flip more than one Mn spin. 
We believe that, in order to account for these experimental results, 
a different physical mechanism which involves other degrees of freedom, possibly
nuclear spins  valence band hole angular momentum, might be
needed.  

In summary, we have proposed the possibility of controlling the magnetization
dynamics  of (III,Mn)V ferromagnetic semiconductors by means of laser pulses. 
Our proposal is based on an optical spin transfer effect in which  angular
momentum is transferred from the laser to the collective spin magnetization by
optically oriented photocarriers.  We have argued that the efficiency in the
spin transfer is close to one in the case of the photo-electrons, but  smaller
in the case of photoholes because of their rapid spin decoherence.  Finally, we
have proposed pump and probe experiments which can achieve all optical FMR and
magnetization switching on ns time scales. 
  
This work has been supported by the Welch Foundation and by the Office of Naval Research under 
grant N000140010951,
MAT2003-08109-C02-01, Ram\'on y Cajal
Program (MCyT, Spain) and UA/GRE03-14.
This work has been partly funded by
FEDER funds. 

\appendix

\section{Derivation of the spin transfer term}
Here we derive an expression for the influence of the photo-carrier
spins on the dynamics of the collective magnetization.  The Mn ion local
moments see the conduction electrons through their mean-field 
interaction with its spin-density.  At each instant 
in time ${\hat \Omega}$ precesses around an effective field 
with a photocarrier contribution $J {\vec s}$: 
\begin{equation}
\frac{{\cal M}_s}{\mu_B}\left.\frac{d\hat{\Omega}}{d t}\right|_{pc}=
\left(\hat{\Omega}\times J \vec{s}\right).
\label{A1}
\end{equation}
The optical spin-transfer torque is specified by the time-dependent spin density
which satisfies
\begin{equation}
\frac{d\vec{s}}{d t}=
-J \hat{\Omega}\times\vec{s} + {\cal P}\hat{n} -\frac{\vec{s}}{\tau}
\end{equation}
Here $J$ is the exchange energy between the quasiparticle and the collective
magnetization, $\tau$ is the photo-carrier spin relaxation time, $\hat{n}$ is
the initial spin polarization of the photo-carrier, and ${\cal P}$ is the
photo-carrier generation rate, as defined in section III.
 We look for solutions of the form:
\begin{equation}
\vec{s}(t)=e^{-t/\tau} \vec{v}(t) + \vec{s}_{0}
\label{ans}
\end{equation}
where $\vec{v}(t)$ satisfies ${\dot \vec{v}}(t)=J\hat{\Omega}\times\vec{v}$.
  It follows that 
\begin{equation}
\vec{s}_0 + \xi \hat{\Omega}\times \vec{s}_{0} = {\cal P} \hat{n}
\label{A2}
\end{equation}
where $\xi\equiv J\tau$  is the dimensionless (taking $\hbar=1$) ratio between 
the photo-carrier decay time and precession period.Equation (\ref{A2}) 
can be written as a matrix equation
$$A\vec{s}_0= \tau {\cal P} \hat{n}$$
 where $A_{ij}=\delta_{ij}+\xi
\epsilon_{ijk}\Omega_k$. $\vec s_{0}=A^{-1} \tau {\cal P} \hat n$
where 
\begin{equation}
A_{ij}^{-1}=\frac{1}{1+\xi^2} \left(\delta_{ij}+\xi^2\Omega_i\Omega_j -\xi
\epsilon_{ijk} \Omega_k \right).
\end{equation} 
Since $\xi>>1$ for conduction band photo-carriers we can drop 
$O(\frac{1}{\xi^2})$ terms to obtain:
\begin{equation}
\vec{s}_{0} =\tau {\cal P}
\left(\hat{\Omega}\cdot\hat{n}\right)\hat{\Omega}+\frac{{\cal P}}{J}.
\hat{n}\times\hat{\Omega}
\label{A3}
\end{equation}
Because of the high precession rate of the photo-carriers, the precessing
contribution  proportional to $\vec{v}(t)e^{-t/\tau}$ is unimportant  even for
$t$ shorter than $\tau$.  Therefore, combining equations
(\ref{A1},\ref{ans},\ref{A2},\ref{A3}) we  find that the contribution of the
photocarriers to the collective magnetization dynamics is
\begin{equation}
\frac{{\cal M}_s}{\mu_B}\left.\frac{d\hat{\Omega}}{d t}\right|_{pc}=
{\cal P}\hat{\Omega}\times\left(\hat{n}\times\hat{\Omega}\right).
\label{A6}
\end{equation}
which leads to eq. (\ref{ST}).
This contribution corresponds to the exchange field from the portion of the injected 
spin-density that is perpendicular to the magnetization.  Note that in equation \ref{A2}
we might have distinquished longitudinal and transverse spin-relaxation rates; this 
distinction would not have mattered in the end, essentially because only the 
transverse component produces a spin torque. 
This analysis leads to the same conclusion as the qualitative discussion in the 
main text, which appears superficially to follow a different line.  The difference
is simply one of bookkeeping.  In this appendix we consider the photocarrier spin density at a 
given point in space and time, instead of following the history of a given 
photocarrier from the moment of generation to spin-decay.

\section{Calculation of the conduction band spin decoherence time}
In this appendix we outline the 
 calculation of the conduction band spin decoherence time $T_2$, which
  is done in
the framework of 'system plus reservoir' master of equation (ME) approach
\cite{Cohen}. The system is the conduction band electron spin and the reservoir
is formed by the spin waves of the Mn-hole system {\and} the orbital degrees of
freedom of the conduction band electron. The Hamiltonian reads:
\begin{eqnarray}
{\cal H}_{\rm cond}&=& {\cal H}_{0s}+
{\cal H}_{0r}+{\cal V}_{sr},\nonumber \\
{\cal H}_{0s}&=&\frac{\Delta_c}{2} \vec{\sigma}\cdot \vec{\cal M}_0
\nonumber \\ 
{\cal H}_{0r}&=&\frac{p^2}{2m}+ \sum_{q}\omega(q)b^{\dagger}_q b^{\dagger}_q
 \nonumber \\ 
 {\cal V}_{sr}&=&\frac{J_{sd}}{2}  \sum_{I} \delta\left(\vec{r}-\vec{R}_I\right) \vec{\sigma}\cdot
\delta\vec{M}(\vec{R}_I)
\label{B1}
\end{eqnarray}
where $\Delta_c=J_{sd} c_M$.  The spin fluctuations 
$\delta\vec{M}(\vec{R}_I)$ and spin wave operators are related by
\cite{Konig}:
\begin{eqnarray}
{\cal M}^z(\vec{R})\equiv S- b^{\dagger}(\vec{R})
b(\vec{R}) \nonumber \\
\delta{\cal M}^{(+)}(\vec{R})= {\cal M}^{(+)}(\vec{R})\simeq \sqrt{2S} b(\vec{R})
\nonumber \\
\delta{\cal M}^{(-)}(\vec{R})=
{\cal M}^{(-)}(\vec{R})\simeq \sqrt{2S} b^{\dagger}(\vec{R})
\end{eqnarray}
and
\begin{eqnarray}
b(\vec{R})\equiv  \frac{1}{\sqrt{{\cal N}}}\sum_{\vec{q}} e^{i
\vec{q}\cdot \vec{R}} b_{\vec q} 
\nonumber \\
b^{\dagger}(\vec{R})\equiv
\frac{1}{\sqrt{{\cal N}}}
\sum_{\vec{q}} e^{i
\vec{q}\cdot \vec{R}} b^{\dagger}_{-\vec q}
\end{eqnarray}

The ${\cal H}_{0s}$ term is the
interaction of the spin with the average magnetization
which, in the master equation approach, is the 'system'
Hamiltonian. ${\cal H}_{0r}$ accounts for the
electron kinetic energy and free  spin waves. They are
the 'reservoir' Hamiltonian. The last term, the
coupling between the system and the reservoir
variables,  comes from the exchange coupling of the
conduction electron with the spin waves (spin
fluctuations).

Using second order perturbation theory around  $ {\cal H}_{0s}+{\cal H}_{0r}$
it is possible to derive \cite{Cohen} a closed set of ME for the reduced
density matrix of the conduction band spin, including the coupling with the
reservoir degrees of freedom to second order in the exchange coupling. In this
language, the above Hamiltonian describes elementary  processes  in which a
spin wave is absorbed or emitted, and  spin and momentum are exchanged between
the spin wave and the photo-carrier. Let us  denote by 
\begin{equation}
\Gamma_{\sigma_i k_i
sw_i,\sigma_f,k_f,sw_f}=\frac{2\pi}{\hbar}|{\cal V}_{if}|^2\delta(E_i-E_f)
\end{equation}  

the Fermi Golden rule (FGR) transition rate for the process
in which the photo carrier spin goes from $\sigma_i$ to $\sigma_f$, the
photocarrier momentum goes from $k_i$ to $k_f$ by emission  or absorption  of a
spin wave. We now define 
\begin{equation}
\Gamma_{\sigma_i,\sigma_f}\equiv\sum_{k_i,sw_i}
P(k_i) P(sw_i) \sum_{kf,sw_f}
\Gamma_{\sigma_i k_i sw_i,\sigma_f,k_f,sw_f} 
\label{B2}
\end{equation}
where $P(k_i)$ and $P(sw_i)$ are the equilibrium distribution functions for 
the inital photocarrier momentum and spin wave occupation 
respectively. Eq.
(\ref{B2}) involves both an average over initial and sum over final 
reservoir states
of the Fermi Golden rule transition rates.  It can be seen \cite{Cohen}
 that the spin decoherence time is
 \begin{equation}
 \frac{1}{T_2}=\Gamma_{\uparrow,\downarrow}+\Gamma_{\downarrow,\uparrow}
 \end{equation}
 since the so called non adiabatic contribution to $T_2$ is zero for Hamiltonian
 {\ref{B1}). After some work, the spin wave emission rate reads:
\begin{eqnarray}
\Gamma_{\uparrow,\downarrow}=\frac{V}{Z} \frac{J_{sd}^2}{\hbar} 2S\pi 
\int_{{\cal V}_D} \frac{d\vec{q}}{(2 \pi)^3}
\int \frac{d{\vec k}}{(2 \pi)^3}
\nonumber \\
\int \frac{d{\vec k}'}{(2 \pi)^3}
e^{-\beta \epsilon_k} \left[1+n_B(\Omega(q))\right] \delta\left[\epsilon_k -
\epsilon_{k'} + \Delta -\Omega \right]
\label{B7}
\end{eqnarray}
and a similar expression can be derived for the spin wave absorption rate. 
Here  $V$ is the volume of the sample,  ${\cal V}_D$ is the Debye sphere
and 
$Z=  \frac{V\sqrt{\pi}}{8 \pi^2} 
\left[\frac{2m k_bT}{\hbar^2}\right]^{(3/2)}  
$.  Equation (\ref{T2}) is obtained from eq. (\ref{B7}) after some extra
changes of variables.

%\narrowtext
%%%%%%%%%%%%%%%%%%%%%%%%%%%%%%%%%%%%%%%%%%%%%%%%%%%%%%%%%%%%%%%%


\begin{references}
\bibitem{note} There is not yet complete consensus on the mechanism responsible for
ferromagnetism or  on the low-energy effective theory in (III,Mn)V materials.  
Although we appeal to a picture in which S=5/2 local moments are coupled by
valence band holes, our main qualitative conclusions do not depend on this
assumption.
\end{references}
\end{document}